\documentclass[12pt]{article}
\pdfoutput=1
\usepackage{graphicx}
\usepackage{epsfig}
\usepackage{epstopdf}
\DeclareGraphicsExtensions{.pdf,.eps,.png,.jpg,.mps}

\usepackage{url}
\usepackage[implicit=false]{hyperref}
\usepackage{etoolbox}
\appto\UrlBreaks{\do\-}
\usepackage{cite}

\usepackage[T1]{fontenc}
\usepackage{lmodern}

\def\lsim{\mathrel{\rlap{\lower3pt\hbox{\hskip0pt$\sim$}}
     \raise1pt\hbox{$<$}}}         
\def\gsim{\mathrel{\rlap{\lower4pt\hbox{\hskip1pt$\sim$}}
     \raise1pt\hbox{$>$}}}         

\usepackage{amsmath}
\usepackage{amsfonts}

\begin{document}
\begin{titlepage}

\centerline{\Large \bf iCurrency?}
\medskip

\centerline{Zura Kakushadze$^\S$$^\dag$\footnote{\, Zura Kakushadze, Ph.D., is the President of Quantigic$^\circledR$ Solutions LLC,
and a Full Professor at Free University of Tbilisi. Email: \href{mailto:zura@quantigic.com}{zura@quantigic.com}} and Willie Yu$^\sharp$\footnote{\, Willie Yu, Ph.D., is a Research Fellow at Duke-NUS Medical School. Email: \href{mailto:willie.yu@duke-nus.edu.sg}{willie.yu@duke-nus.edu.sg}}}
\bigskip

\centerline{\em $^\S$ Quantigic$^\circledR$ Solutions LLC}
\centerline{\em 1127 High Ridge Road \#135, Stamford, CT 06905\,\,\footnote{\, DISCLAIMER: This address is used by the corresponding author for no
purpose other than to indicate his professional affiliation as is customary in
publications. In particular, the contents of this paper
are not intended as an investment, legal, tax or any other such advice,
and in no way represent views of Quantigic$^\circledR$ Solutions LLC,
the website \url{www.quantigic.com} or any of their other affiliates.
}}
\centerline{\em $^\dag$ Free University of Tbilisi, Business School \& School of Physics}
\centerline{\em 240, David Agmashenebeli Alley, Tbilisi, 0159, Georgia}
\centerline{\em $^\sharp$ Centre for Computational Biology, Duke-NUS Medical School}
\centerline{\em 8 College Road, Singapore 169857}
\medskip
\centerline{(August 28, 2019)}

\bigskip
\medskip

\begin{abstract}
{}We discuss the idea of a purely algorithmic universal world iCurrency set forth in \cite{iGDP} and expanded in \cite{CryptoRub} in light of recent developments, including Libra. Is Libra a contender to become iCurrency? Among other things, we analyze the Libra proposal, including the stability and volatility aspects, and discuss various issues that must be addressed. For instance, one cannot expect a cryptocurrency such as Libra to trade in a narrow band without a robust monetary policy. The presentation in the main text of the paper is intentionally nontechnical. It is followed by an extensive appendix with a mathematical description of the dynamics of (crypto)currency exchange rates in target zones, mechanisms for keeping the exchange rate from breaching the band, the role of volatility, etc.
\end{abstract}
\bigskip
\medskip

\noindent{}{\bf Keywords:} cryptocurrency, universal numeraire, universal world currency, supply and demand, Libra, volatility, stability, FX, target zone, attainable boundaries, Brownian motion, drift, reflecting barriers, government, HKD, USD, EUR, JPY, monetary authority, currency board, reserve, Bitcoin, stablecoin, blockchain, liquidity, trading, finance, capital markets, cryptoasset, currency basket, carry trade, interest rate
\end{titlepage}

\newpage


{}The idea of a universal world currency is not new.\footnote{\, For some literature on this and related topics, see, e.g., \cite{Bordo2006}, \cite{Cooper1984}, \cite{Cooper2006}, \cite{Friedman1951}, \cite{Graham1944}, \cite{Graham1941}, \cite{Hart1976}, \cite{Hayek1943}, \cite{Iwamoto1997}, \cite{Kaldor1964}, \cite{Keynes1943}, \cite{Mundell2005}, \cite{Rogoff2001}, \cite{Schmukler2006}, \cite{Starr2006}, \cite{Steil2007}, \cite{Ussher2009}, \cite{Welfens2006}, and references therein.} With the advent of Bitcoin \cite{Nakamoto} and other cryptocurrencies,\footnote{\, Also see, e.g., \cite{Berentsen2018}, \cite{Buterin2014}, \cite{CryptoRub} (for a host of references), \cite{Schwartz2014}, \cite{White2015}.} this idea has acquired a brand new angle. Thus, in December 2014 the paper \cite{iGDP}\footnote{\, This paper was uploaded to SSRN on Dec 24, 2014 and officially processed by SSRN on Dec 26, 2014. The full PDF of this paper was freely downloadable from SSRN until the paper was published in {\em The Journal of Portfolio Management} (JPM) in Spring 2015 as an Invited Editorial. Its full PDF was freely downloadable from JPM's website until circa early 2018, when JPM's website was migrated, and with 12,000+ downloads this paper was JPM's most downloaded article of all time. So, apparently, the paper had a wide distribution and its contents were widely known.} was posted on SSRN,\footnote{\, Social Science Research Network (\url{https://www.ssrn.com/index.cfm/en/}).} and this paper proposes a purely algorithmic universal world {\em iCurrency}.\footnote{\, It is unknown what ``i'' stands for in iCurrency. It has nothing to do with iPhone though.}

{}The paper \cite{iGDP} set forth the following criteria for iCurrency: 1) it is not a currency issued (or backed) by any government; 2) it is valued based solely on supply and demand; 3) it is easily transferred across regions and globally accepted as a payment method; and 4) it is algorithmic, with no human intervention. At that time, this might have seemed (and probably did) like a farfetched utopian idea. E.g., based on the wild historical volatility of Bitcoin and other cryptocurrencies, criterion 2) above may appear to be most challenging in terms of stability of iCurrency, which would be important for universal adoption.

{}Subsequently, the paper \cite{CryptoRub}\footnote{\, This paper was uploaded to SSRN on Oct 25, 2017 and officially processed by SSRN on Oct 26, 2017, with subsequent minor revisions focused on anti-money laundering (AML), sustainability (electricity consumption due to mining) and effects on bank fees of government-issued cryptocurrencies (to appeal to the readership base of {\em Risk}, where a shortened version of this paper was published under the title ``Russian crypto-currency will threaten AML efforts''), but with the iCurrency section unaltered. Importantly, the full PDF of the paper has been freely downloadable from SSRN since Oct 25, 2017, including the consortium and target zone ideas (see below). Furthermore, the paper was discussed by the press \cite{Bershidsky2017}, \cite{Holmes2017}, with other outlets reprinting these articles.} expanded on the ideas of \cite{iGDP} with two important new insights. First, iCurrency could be issued by a broad {\em consortium} of governments, thereby preserving criterion 1) above.\footnote{\, As well as aiding in satisfying criterion 3) above.} The idea here is very simple. To avoid any given government or a group of governments going rogue and manipulating iCurrency,\footnote{\, E.g., via (analogs of) central banks artificially increasing or decreasing interest rates, monetary supply manipulation, political influences, and so forth.} all is required is that a network collusion (of the majority or super-majority of the network members, depending on the precise protocol) is virtually impossible. This does not require an enormous number of miners -- unlike decentralized Bitcoin, with a consortium, there is no need for mining, and about 100 or so consortium members would suffice.\footnote{\, So long as the protocol is 100\% democratic, unlike, e.g., the UN Security Council protocol.}

{}Second, \cite{CryptoRub} further proposes that, to achieve stability and low volatility in the context of criterion 2) above, a staged rollout of iCurrency may be warranted, whereby its value is determined by a combination of supply and demand and some regulation by a central authority, e.g., via a target zone similar to USD/HKD, which trades between 7.75 and 7.85, a band fixed by the Hong Kong Monetary Authority (HKMA). So, iCurrency too could be regulated at first to stay in such a target zone, which could be gradually relaxed and perhaps removed entirely at some future time.\footnote{\, ``Stability'' and ``volatility'' are meaningful only if there is another numeraire to compare a currency with. If iCurrency were the only currency in the entire world, then it would appear to have no choice but to be stable. What could be unstable is prices of goods and services in some parts of the world (inflation). However, this would not make iCurrency unstable per se. Indeed, hyperinflation in, e.g., Zimbabwe or Venezuela does not destabilize USD, EUR or JPY.\label{fn.stability.vol}} Or perhaps some wide target zone could still be maintained.\footnote{\, In which case iCurrency would not be 100\% based on {\em unregulated} supply and demand (see below), but could be close to it for all practical purposes, and the band would ensure some nominal degree of stability in case of unforeseen impactful events.}

{}To be clear, even in the HKD/USD-style target zone iCurrency would still be valued based solely on supply and demand as the regulation by a central authority itself is based on supply and demand. E.g., in the case of USD/HKD, HKD is fully backed by USD reserves, and HKMA, which is a currency board, intervenes when HKD is weak (i.e., hits the upper boundary of 7.85 of the target zone) or strong (i.e., hits the lower boundary of 7.75 of the target zone) by buying HKD in exchange for USD (thereby reducing the HKD supply) or selling HKD in exchange for USD (thereby increasing the HKD supply), respectively. So, because HKMA does not create HKD out of thin air and HKD is fully backed by USD reserves, USD/HKD is 100\% based on supply and demand (just as in criterion 2) above for iCurrency), except that some of that supply and demand is unregulated (that stemming from free market participants) and some of it is regulated (that stemming from HKMA).\footnote{\, The HK\$10 notes and coins are issued by the Hong Kong Government. Other notes are issued by three note issuing banks, to wit, The Hongkong and Shanghai Banking Corporation Limited, the Bank of China (Hong Kong) Limited, and the Standard Chartered Bank (Hong Kong) Limited. See, e.g., \url{https://www.hkma.gov.hk/eng/key-functions/monetary-stability/notes-coins-hong-kong/notes.shtml}.}

{}Another important aspect of Hong Kong's Linked Exchange Rate System (LERS) \cite{HKMA2011} is that Hong Kong interbank interest rates closely follow their U.S. counterparts; the Hong Kong short-term interest rates are essentially ``pegged'' to the U.S. short-term interest rates, with the spread between them reflecting the premium/discount stemming from the USD/HKD exchange rate. Thus, under Uncovered Interest Rate Parity (UIRP),\footnote{\, For some literature on UIRP and related topics, see, e.g., \cite{Anker1999}, \cite{Ayuso1996}, \cite{Bacchetta2006}, \cite{Baillie2000}, \cite{Bekaert2007}, \cite{Beyaert2007}, \cite{Bilson1981}, \cite{Chaboud2005}, \cite{Engel1996}, \cite{Fama1984}, \cite{Frachot1996}, \cite{Froot1990}, \cite{Hansen1980}, \cite{Harvey2015}, \cite{Hodrick1987}, \cite{Ilut2012}, \cite{Lewis1995}, \cite{Lustig2007}, \cite{Mark2001}, \cite{Roll2008}.\label{fn.UIRP}} if HKD is weak (i.e., USD trades above the midpoint 7.80 of the target zone, closer to 7.85), then the Hong Kong interest rates should be lower than their U.S. counterparts. Similarly, if HKD is strong (i.e., USD trades below the midpoint 7.80 of the target zone, closer to 7.75), then the Hong Kong interest rates should be higher than their U.S. counterparts. Otherwise there will be arbitrage opportunities through carry trades (see Appendix \ref{sec2} for details).\footnote{\, However, as we discuss in Appendix \ref{sec2}, pegging interest rates ``perfectly'', that is, to eliminate such arbitrage opportunities, may not always be possible in high FX rate volatility environments. Also, note that deviations from UIRP are not risk-free arbitrage opportunities (see below).}

{}In fact, HKMA's role in ensuring that USD/HKD does not breach the band cannot be underestimated. Thus, according to its 2018 report \cite{HKMA2019}, HKMA intervened 27 times and bought HK\$103.5 billion ``from the market in an orderly and transparent manner'' to counteract the weakening of HKD against USD at the upper (7.85) boundary. According to \cite{HKMA2019}, this weakening was caused by rising U.S. interest rates, which ``attracted increased interest carry trade activities involving the selling of HKD in exchange for USD''. The bottom line is that, without HKMA's aforesaid serial interventions, USD/HKD would invariably breach the upper boundary of the band in 2018. Furthermore, note that the HKMA's monetary policy intervention process ``is very much an automatic mechanism'' \cite{HKMA2011}.

{}Hong Kong's LERS has been in place for over 35 years and has withstood the test of time. In this regard, the proposal of \cite{CryptoRub} to use an USD/HKD-style target zone to achieve iCurrency stability appears to make sense. In fact, a recent development -- the proposed Libra cryptocurrency \cite{Libra2019} -- has put yet another twist on the idea of iCurrency. Libra, if materialized,\footnote{\, There are many hurdles for Libra, including regulatory ones (see, e.g., \cite{Kihara2019}, \cite{Marsh2019}, \cite{USHCFS2019}). However, there are also other, more basic issues that must be addressed (see below). For critiques of the Libra proposal from other angles, see, e.g., \cite{Fatas2019}, \cite{White2019}. Also, for the technical details of the Libra blockchain, see \cite{Amsden2019}.} a priori could be a contender for iCurrency. In fact, two of the main ideas behind Libra are precisely those set forth in \cite{CryptoRub}, to wit, i) Libra is governed by the independent Libra Association \cite{Libra2019}, which is the {\em consortium} proposed in \cite{CryptoRub}, and ii) Libra is operated similarly to ``currency boards (e.g., of Hong Kong)'' and coins can be converted ``back to fiat at a narrow spread above or below their current value'' \cite{Catalini2019}, which is the USD/HKD-style {\em target zone} proposed in \cite{CryptoRub}.\footnote{\, Recently a perplexing article \cite{Allison2019} was published with the title ``MIT Fellow Says Facebook `Lifted' His Ideas for Libra Cryptocurrency'', the claim apparently being that those ideas were ``lifted'' from \cite{Lipton2018}. Both \cite{iGDP} and \cite{CryptoRub} substantially predate \cite{Lipton2018}, so the claim that Facebook ``lifted'' the ideas for Libra from \cite{Lipton2018} has no leg to stand on. If any papers deserve credit, those would appear to be \cite{iGDP} and \cite{CryptoRub}. To be clear, we have no quarrel with Libra: strictly speaking, commercially-oriented white papers (which are not academic papers or patent applications) are not required to cite others, albeit, unsurprisingly, it would certainly be pleasant if the Libra white papers \cite{Libra2019} and \cite{Catalini2019} cited \cite{iGDP} and \cite{CryptoRub}. On the other hand, there appears to be no excuse for \cite{Lipton2018} (an academic journal article) not citing \cite{iGDP} and \cite{CryptoRub}, especially considering that one of the authors of \cite{Lipton2018} had an email correspondence about \cite{CryptoRub} with one of the authors of \cite{CryptoRub} in January 2018.} So, could Libra possibly become iCurrency?

{}There are some issues with the Libra proposal in its current form (see \cite{Libra2019}, \cite{Catalini2019}). First, Libra will be backed by ``a collection of low-volatility assets, including bank deposits and government securities in currencies from stable and reputable central banks'' (i.e., high credit rating assets to ensure capital preservation), and these will be short-maturity assets to guarantee liquidity \cite{Catalini2019}. So, for all intents and purposes, ``the value of Libra will be effectively linked to a basket of fiat currencies'' \cite{Catalini2019}. At first, it might appear that, backing Libra by a reserve (effectively) holding a diverse basket of fiat currencies (as opposed to one currency as in the case of HKD) would help make it more stable and less volatile. However, as mentioned above, ``stability'' and ``volatility'' are meaningful only if there is another numeraire to compare a currency with (see fn. \ref{fn.stability.vol}). I.e., the question is w.r.t. what numeraire would Libra be ``more stable''/``less volatile'' if it is (effectively) backed by a basket of fiat currencies? Is this numeraire USD, gold, a basket of commodities, or...? I.e., what is the yardstick?

{}The issue here is that currently there is no supranational/universal numeraire that iCurrency is meant to be. So, if Libra's stability is measured by using USD as the numeraire, then there is a hiccup here. Uniform weighting is not critical here, so for the sake of simplicity let us assume that the underlying fiat basket is uniformly weighted (i.e., all $N$ fiat currencies in the reserve basket initially have the same USD amounts).\footnote{\, The reserve basket is not intended to be actively managed \cite{Catalini2019}.} If $N$ is sizable (e.g., $N \geq 10$), then the USD contribution is relatively small and the basket is dominated by the $(N-1)$ foreign currencies. However, this is simply a bet against the state of the U.S. economy. Indeed, the essence of the so-called dollar carry trade (see, e.g., \cite{Lustig2014}) is to be long (short) USD and short (long) a diversified basket of foreign currencies (typically, with equal weights) when the average cross-sectional forward discount\footnote{\, The forward discount for a given currency is defined as the natural logarithm of the ratio of the spot FX rate over the forward FX rate.} for the foreign currency basket is negative (positive). Empirical evidence suggests that this strategy relates to the state of the U.S. economy: when the U.S. economy is weak (strong), the average forward discount tends to be positive (negative).\footnote{\, See, e.g., \cite{Cooper2008}, \cite{Joslin2018}, \cite{Joslin2014}, \cite{Lustig2014}, \cite{Stambaugh1988}, \cite{Tille2001}.} A bet against the U.S. economy hardly would make much sense as a reserve backing Libra. One way to mitigate this is to have nonuniform weights with a large weight allocated to USD (with EUR and JPY also having larger weights than other currencies in the basket). However, this actually does not address the main problem, to wit, that ``stability'' is only meaningful w.r.t. a given numeraire, which usually defaults to USD, the de facto stable currency of choice in countries with unstable currencies.\footnote{\ This is what \cite{HKMA2011} states regarding possibly linking HKD to a basket of currencies (instead of USD): ``By linking to a basket of currencies, the domestic economy would be less
exposed to sharp swings in the exchange rate and interest rates of a single anchor currency. But the system would be more complex and much less transparent. To the extent that the monetary authority retained discretion to adjust the weights of the component currencies, transparency and predictability would be reduced, possibly undermining confidence in the exchange rate system.''}

{}Based on the foregoing, the idea of a broad currency basket,\footnote{\, According to \cite{Libra2019}, ``The assets in the Libra Reserve will be held by a geographically distributed network of custodians with investment-grade credit rating to provide both security and decentralization of the assets.'' While ``decentralization'' might appear appealing, if (an equivalent of) a diverse basket of currencies is to be held in the reserve, it might be much simpler, more efficient and less costly to use ETFs to achieve this. E.g., Invesco DB G10 Currency Harvest Fund (ticker DBV) tracks Deutsche Bank G10 Currency Future Harvest Index -- Excess Return (note that DBV currently has only about \$24M in the assets under management (AUM)); SPDR Barclays Capital Short Term International Treasury Bond ETF (ticker BWZ) tracks Barclays Capital 1-3 Year Global Treasury ex-US Capped Index (BWZ currently has about \$293M in AUM); iShares 1-3 Year International Treasury Bond ETF (ticker ISHG) tracks S\&P/Citigroup International Treasury Bond Index Ex-US 1-3 Year (ISHG currently has about \$66M in AUM). There are additional ETFs based on all-duration developed country treasury bond baskets, e.g.: SPDR Barclays International Treasury Bond (ticker BWX) tracks Barclays Capital Global Treasury Ex-US Capped Index (BWX currently has about \$1.14B in AUM); iShares International Treasury Bond ETF (ticker IGOV) tracks S\&P/Citigroup International Treasury Bond Index Ex-US (IGOV currently has about \$963 AUM); plus lower AUM ETFs (ticker IGVT with about \$11M in AUM, and ticker FLIA with about \$5M in AUM). There is a host of U.S. Treasury ETFs. In addition, there are currency-specific ETFs, including USD-specific ETFs such as USD-bullish ETFs (ticker UUP with about \$346M in AUM, and ticker USDU with about \$43M in AUM), and a USD-bearish ETF (ticker UDN with about \$37M in AUM and down about 2.57\% YTD, whose constituents are EUR, JPY, GPB, CAD, SEK and CHF). Depending on the reserve objectives, there are other ETFs that may also be suitable. However, once again, (an equivalent of) a diversified currency basket may not be as desirable as it might appear at first (see above).} as appealing as it may sound at first, apparently is too simplistic a solution and by no means a panacea. In fact, the idea of using a currency basket (e.g., a basket of USD, EUR and JPY) is not new and has its own issues (see, e.g., \cite{Cooper2006}, \cite{Schmukler2006}, \cite{Welfens2006}). A diversified basket does not eliminate fluctuations of individual currencies, so the prices of goods in Libra in any given country will fluctuate with its fiat currency (and this is recognized in \cite{Libra2019}, \cite{Catalini2019}). One of the hurdles to broad adoption is precisely this simple fact, that if there are many free-floating fiat currencies around, such fluctuations in the prices of goods is difficult to avoid. Thus, contrary to early unrealized expectations/predictions \cite{Friedman1953}, empirically FX rates fluctuate much more than the prices of goods,\footnote{\, For some literature on this and related topics, see, e.g., \cite{Baxter1989}, \cite{Frankel1993}, \cite{Goldberg1988}, \cite{Krugman1991}, \cite{Obstfeld2000}, \cite{Rogoff2001}, and references therein.} and there is little that can be done about this apart from implementing a USD/HKD-style target zone and achieving stability of prices in the local currency (HKD) w.r.t. the reference currency (USD). This, however, cannot achieve stability w.r.t. the entire plethora of various other currencies. {\em As\'{i} es la vida.}

{}Another issue with the Libra proposal is that, while the reserve collects the interest paid by the underlying securities purchased with the fiat funds received from the Libra holders, the latter do not receive any interest on their Libra holdings at all \cite{Libra2019}, \cite{Catalini2019}. As such, this is nothing but a vanilla carry trade, where the Libra Association members earn interest,\footnote{\, In this regard, if the reserve is based on a diverse basket of short-term sovereign debt instruments (as \cite{Libra2019}, \cite{Catalini2019} appear to propose), some of these instruments might not even pay interest in the context of the negative interest rate environments in various jurisdictions. For some literature on negative rates and related topics, see, e.g., \cite{Arteta2017}, \cite{Bech2016}, \cite{Buiter2009}, \cite{Jobst2016}, \cite{Kimball2015}, \cite{Lopez2018}, \cite{Ullersma2002}, \cite{vanRiet2017}, \cite{Yates2004}, and references therein.} while Libra holders do not. This is just pure arbitrage. It is unclear why large institutional players would want to be arbitraged like this. The premise is that certain ``underbanked'' retail users (presumably, with a large fraction coming from developing countries) would find value in holding or transacting in Libra, and that is supposed to be a justification enough to want to be arbitraged. Larger individual or institutional players may wish to subject themselves to such arbitrage if their intents are nefarious (money laundering, tax evasion, etc.). Either way, it is difficult to see how any of the foregoing adds to the credibility or general appeal of the Libra proposal. If shorting Libra is not the sole privilege of the Libra Association members (if it is, then it means the Libra Association members just want to arbitrage the rest of the world...) and if it is possible to do such shorting at a reasonably low cost (including via forward contracts, futures, ETFs, options and other vehicles), then there would be an arbitrage opportunity for others to execute this carry trade putting possibly substantial selling pressure on Libra. It is difficult to see how this can be avoided (again, unless the general public is restricted from shorting Libra).

{}Now we come to the most pressing issue with the Libra proposal: ``The association does not set monetary policy... Since Libra will be global, the association
decided not to develop its own monetary policy but to inherit the policies of the central banks represented in the basket'' \cite{Catalini2019}. This appears to implicitly assume that Libra is expected to remain stable and trade in a narrow band solely based on the fact that i) it is backed by a reserve, and ii) ``the Libra Reserve acts as a ``buyer of last resort''{}'' \cite{Libra2019}. It appears that, in this regard, the Libra proposal (implicitly) assumes that: i) if Libra trades (substantially) lower than its ``fair'' (i.e., ``reserve-based'') value, then the market will buy Libra and push its price higher; and ii) if Libra trades (substantially) higher than its aforesaid ``fair'' value, then the Libra Association members will sell more Libra coins to the market and push its price lower. However, this is just way too simplistic. Just because Libra is backed by a reserve, it does not guarantee that such ``arbitrage'' argument is valid, as it does not take into account various important aspects, including irrationality of the market. On the weak-side, there is no guarantee that the market will want to
buy more Libra just because it is ``cheap''. On the strong-side, there is no guarantee that the Libra Association members will place more of their fiat funds into the reserve to mint new Libra coins to sell to the market: not only is there risk involved in this, but there could be liquidity issues as the Libra Association members do not have unlimited resources (see below) and unlike central banks cannot create money out of thin air. Furthermore, unless the Libra Association members are {\em obligated} to do so, expecting them to do this on their own, without a stringent mechanism in place, is simply too generous an assumption.

{}We do not have to go far to see that these are serious issues. Thus, HKD is fully backed by a USD reserve. In fact, as of the end of July 2019, the HKMA Exchange Fund balance was HK\$3,686.2 billion, while the Monetary Base was HK\$1,633.1 billion (see \url{https://www.hkma.gov.hk/eng/key-information/press-releases/2019/20190814-3.shtml}). The Backing Ratio, defined as the Backing Assets (a part of the Exchange Fund) divided by the Monetary Base (comprising Certificates of Indebtedness, Government issued currency notes and coins in circulation, the balance of the banking system and Exchange Fund Bills and Notes issued) was 109.9\% on Dec 31, 2018 \cite{HKMA2019}. Yet, as discussed above, HKMA had to intervene whopping 27 times in 2018 on the weak-side to prevent HKD from breaching the band (see above). Backing a currency by a reserve is simply insufficient for keeping its exchange rate in a target zone. Volatility can be caused by rational as well as irrational market participants, reactions to the news, etc. Furthermore, as discussed above, FX rates fluctuate much more than the prices of goods. Thus, one can take an ``asset'' view of exchange rates, whereby a currency is a durable and its price reflects {\em future} flows, not just its transactions value at a given time, so it is not surprising that FX rates fluctuate so much (cf. stocks). And the feedback of these fluctuations into the economy is damped (among other things) by trade costs, including tariffs, transport costs, costs related to differences in languages and legal systems, etc. (see, e.g., \cite{Rogoff2001} and references therein). Fiat currencies of large democracies are backed by their governments (that wield their taxing power, etc.). Yet, their currencies fluctuate. HKD is fully backed by USD, yet it fluctuates and is kept inside the target zone by HKMA's monetary interventions, not just the fact that it is backed by USD. Fundamentally there is no reason why Libra would be any more stable than many (developed countries') fiat currencies.\footnote{\, If anything, a priori it could be expected to be less stable. Thus, notwithstanding stratospheric market capitalizations of some of Libra's backers, these are not liquid funds. In fact, the initial contributions of the Libra Association members are only \$10M each \cite{Libra2019}, which even with the target number of 100 members is only \$1B. With the ambition to become a global currency, it would take a much, much larger reserve to sustain Libra. Is this reserve supposed to be raised from Libra users? Then how is this any different from many of the ubiquitous cryptocurrency, digital coin and token projects designed to raise money from the public with no tangible prospects of returns? Furthermore, if this is the case, then it only makes the optics that this is an attempt at arbitraging the rest of the world through a carry trade (see above) even worse.} If the market perceives that there is a potential {\em future} problem with Libra's backing, its reserve, any of the major members of the Libra Association, etc., the market will not hesitate to sell, sell, sell. Similarly, if the market believes that Libra will have {\em future} value beyond what it is backed with, the market will not hesitate to buy, buy, buy. Bitcoin is not backed by any reserve and on the face of it has no intrinsic value, but this has not stopped the market -- whether rationally or irrationally -- from trading Bitcoin in an unfathomably wide range. Backing Libra with a reserve but without a well-defined monetary policy will not prevent its price from fluctuating wildly.\footnote{\, Consider this. As a {\em Gedankenexperiment} (thought experiment), imagine that -- magically -- starting tomorrow Bitcoin is backed by USD reserves such that each BTC is backed by one USD. Does this mean that the BTC price will immediately drop to 1 USD? Of course not! Such backing might introduce some volatility in the short term (then again, BTC is already extremely volatile), but the market will continue to trade BTC at values it {\em perceives} to be ``right'' at any given time.}

{}To reiterate, being backed by other assets does not confine a currency into a narrow band. The only relevant ``arbitrage'' argument is that of UIRP (and not that a currency is backed by a reserve), and even that is not failproof as deviations from UIRP occur all the time (see, e.g., references in fn. \ref{fn.UIRP}). In terms of UIRP, the ``no-arbitrage'' argument goes as follows (note that deviations from UIRP do not give rise to risk-free arbitrage). The expected rate of return of a cash bond in the domestic currency (e.g., USD), when expressed in the units of the foreign currency (e.g., HKD), must be the same as the expected rate of return of the same investment in a cash bond in the foreign currency, or else an arbitrage opportunity occurs via a vanilla carry trade (by borrowing the currency with a lower rate of return and lending the currency with a higher rate of return). In the context of a target zone this translates into an interest rate ``peg'' between the band-confined currency (HKD) and the free-floating reference currency (USD). And this peg is instituted by the monetary authority of the confined currency (e.g., HKMA in the case of HKD). However, as we discuss in detail in Appendix A, the ideal peg is feasible only in low-volatility environments. If the FX rate volatility is high (notwithstanding the band, i.e., if the FX rate fluctuates with a high frequency within the band), the ideal peg dictated by UIRP would require unrealistically high interest rates on the strong side, and (possibly large) negative interest rates on the weak side. Intuitively, this can be understood as being due to the fact that the interest rate peg basically has to reflect the risk premium due to high volatility. In realistic high-volatility scenarios (and there is no reason to expect Libra not to fit this profile) the limited peg that can be realistically instituted invariably allows carry trade arbitrage, so the band can be breached.\footnote{\, As mentioned above, according to \cite{Libra2019}, the interest earned on Libra is simply zero. Then, unless only the Libra Association members are allowed to short Libra and other market participants are precluded from doing so -- that would be a tough pill to swallow -- there will be carry trade arbitrage opportunities (see above) putting downward pressure on Libra.} This is precisely what happened with HKD in 2018 \cite{HKMA2019}, where rising U.S. interest rates put pressure on and weakened HKD to the point that HKMA had to intervene 27 times (see above) to prevent it from breaching the band. The role of the monetary authority and its interventions through a robust monetary policy cannot be underestimated. Without such a monetary policy and interventions, there is no reason to believe that Libra will stay in a narrow band.

{}A lot of attention has gone to regulatory hurdles with Libra. However, even if the regulatory issues magically went away overnight, as it sits, the Libra proposal appears to be too simplistic to work. The consortium idea is sound -- as mentioned above, in the context of iCurrency, this was proposed in \cite{CryptoRub}. The idea of backing a currency with traditional assets is not new and is sound -- HKD is a testament to this -- albeit, as discussed above, backing Libra with a reserve (effectively comprised) of a diverse basket of fiat  currencies is not without potential drawbacks (see above). The idea of having a currency trade in a narrow target zone is also sound -- in the context of iCurrency, this was also proposed in \cite{CryptoRub} -- however, the apparent lack of a monetary policy to keep Libra confined to a band in the Libra proposal simply will not work (see above). The bottom line is that, over many decades, a lot of very smart people (including Nobel laureates) have thought about these and related issues extensively, and expecting what appears to be a rather simplistic proposal to cut it is a bit na\"{i}ve.

{}Libra is an interesting idea in the sense that, if it were to succeed at a global scale, it might be a contender for iCurrency that is backed by a consortium of private entities rather than governments (which is the proposal of \cite{CryptoRub}). The consortium idea, be it comprised of private entities or governments as its members, is appealing as it eliminates the need for wasteful (in terms of energy consumption) mining (as in, e.g., Bitcoin).\footnote{\, For some literature on inefficiency and unsustainability issues, see, e.g, \cite{Bariviera2017}, \cite{Kostakis2014}, \cite{Nadarajah2017}, \cite{Urquhart2016}, \cite{Vranken2017}.} In some sense, the HKD system (LERS -- see above) is a quasi-consortium system as the HKD notes are issued by three commercial banks along with the Hong Kong Government (see above), albeit the monetary policy is implemented solely by HKMA (and not the banks). However, as mentioned above, some aspects of the Libra proposal appear to have shortcomings.

{}In terms of implementing a target zone for iCurrency (and by this we do not mean Libra),\footnote{\, For the sake of completeness, let us mention that stabilization in the context of iCurrency is not to be confused with the so-called stablecoins, which have had their own host of issues (see, e.g., \cite{Chung2018}, \cite{Irrera2018}, \cite{Lian2018}, \cite{Popper2018}, \cite{Robinson2018}). For some literature on stablecoins, see, e.g., \cite{Griffin2018}, \cite{Mita2019}, \cite{Pernice2019}, \cite{Saito2018}, \cite{Senner2018}, \cite{Zhang2019}.} it would appear to make little sense to try to reinvent the wheel. The HKD system (LERS) has worked well for over 35 years and weathered various precarious situations such as the Hong Kong stock market crash of 1987, the Asian financial crisis of 1997/98, the global financial crisis of 2008, etc. Clearly, it works. Therefore, it would make sense to implement the same or very similar system for iCurrency. The process of monetary policy intervention at the boundaries of the target zone ``is very much an automatic mechanism'' \cite{HKMA2011}. So it can be made 100\% algorithmic to comply with criterion 4) for iCurrency (see the four criteria for iCurrency at the beginning of this paper). And, notwithstanding expected political and other issues, it is natural to expect that it would be easier to overcome regulatory hurdles if a broad consortium of governments issues iCurrency as opposed to private entities, albeit if the governments do not or cannot get their act together, an attempt by a private consortium is certainly a welcome development. In fact, as this paper was being finalized, multiple outlets\footnote{\, See, e.g., \cite{De2019}, \cite{Inman2019}, \cite{Lewis2019}.} reported that, in his speech \cite{Carney2019} at the Jackson Hole Symposium 2019, the Governor of the Bank of England Mark Carney floated the idea of iCurrency (which he referred to as ``Synthetic Hegemonic Currency'')\footnote{\, iCurrency sounds so much better!} issued by a network of central banks. So, is the writing on the wall?\footnote{\, As pointed out in \cite{CryptoRub}, government-issued cryptocurrencies may be a step forward (not backward) toward iCurrency. Russia's CryptoRuble (see \cite{CryptoRub}) has not materialized (yet). However, there have been rumors of Chinese government-issued cryptocurrency being ready to be released (possibly soon), see, e.g., \cite{delCastillo2019}.
}

{}Quoting from \cite{iGDP}: ``If mankind is destined to make it to Mars (and
beyond) and establish extraterrestrial colonies in our solar
system, a universal currency devoid of government control
and other forms of manipulation would appear to be
a necessity, rather than wishful thinking. Therefore, we
propose our criteria 1-4 above as the requirements not
only for a universal numeraire for iGDP, but also for any
universal world currency that mankind is very likely destined
to adapt, considering the global nature of the world
economy and our otherworldly aspirations. Cryptocurrencies,
as pioneers of algorithmic and digital currency, might
be positioned to fill this role, perhaps in the not-so-distant
future. Only time will tell.'' It looks like these seemingly utopian ideas might actually be coming to life!

{}Our discussion above is purposefully kept nontechnical to appeal to a broad readership base. However, various important conclusions we reach above (including an essential inference that monetary policy interventions are paramount to keeping the exchange rate from breaching the band) are supported by (not-so-elementary) math, which we present in Appendix \ref{sec2}, parts of which are based on a prior (also rich with math) paper \cite{FXTZ} on FX options in target zones.

\appendix
\section{FX in Target Zone}\label{sec2}

{}The first two subsections of this section follow a condensed version of the discussion in \cite{FXTZ}, which contains more detail on FX target zones and a number of references (which we do not repeat here).

\subsection{Differential Rate}

{}Let us assume that the domestic currency (e.g., USD) is freely traded with no restrictions, whereas the foreign currency (e.g., HKD or a cryptocurrency) trades inside a target zone. We have a domestic cash bond $B^d_t$ and a foreign cash bond $B^f_t$. We also have the exchange rate $S_t$, which, for our purposes here, is the worth of one unit of the domestic currency in terms of the foreign currency (e.g., in our USD/HKD example, $S_t$ is the HKD worth of 1 USD, whose target zone of 7.75 to 7.85 is set by the Hong Kong Monetary Authority (HKMA), Hong Kong's currency board). We will refer to tradables denominated in the foreign (domestic) currency as foreign (domestic) tradables. Say we are interested in pricing derivatives (e.g., FX options, bonds or other instruments) from a {\em foreign} investor's perspective. The foreign cash bond $B^f_t$ is a foreign tradable; however, $B^d_t$ and $S_t$ are not. We can construct another foreign tradable via\footnote{\, In our USD/HKD example, this is the HKD value of the USD cash bond.}
\begin{equation}\label{S-tilde}
 {\widetilde S}_t = B^d_t S_t
\end{equation}
The discounted process, which must be a martingale under the risk-neutral measure, call it ${\bf Q}$, is given by
\begin{equation}\label{ZX}
 Z_t = (B^f_t)^{-1}{\widetilde S}_t = B_t^{-1} S_t
\end{equation}
where
\begin{equation}\label{eff.cash}
 B_t = B^f_t / B^d_t
\end{equation}
The price of a claim $Y_T$ at time $T$ is given by (${\cal F}_t$ is the filtration (or the history) up to time $t$, and $\mathbb{E}(\cdot)$ denotes expectation)
\begin{equation}
 {\widetilde V}_t = B^f_t~\mathbb{E}\left((B^f_T)^{-1} Y_T\right)_{{\bf Q},{\cal F}_t}
\end{equation}
The foreign monetary authority, which confines the foreign currency to the target zone, (in theory) also adjusts (``pegs'') the foreign interest rates based on the domestic interest rates and the FX rate. Therefore, we can assume that the domestic cash bond $B_t^f$ is deterministic within the (short enough) time horizons we are interested in for the purpose of pricing FX derivatives.\footnote{\, More generally, we can assume that any volatility in the domestic bond $B_t^d$ is uncorrelated with the volatility in the FX rate $S_t$ and the volatility in the foreign bond $B_t^f$, or, more precisely, any such correlation is negligible at relevant time horizons. This would not alter any of the subsequent discussions or conclusions, so for the sake of simplicity we will assume that $B_t^d$ is deterministic.}  For the claim price we then have
\begin{eqnarray}
 &&{\widetilde V}_t = B_t^d(B_T^d)^{-1}V_t\label{tildeV}\\
 &&V_t = B_t~\mathbb{E}\left(B_T^{-1} Y_T\right)_{{\bf Q},{\cal F}_t}\label{V_t}
\end{eqnarray}
Note that $B_t$ defined in (\ref{eff.cash}) is the ratio of the two cash bonds. We can define the corresponding differential (or ``effective'') short-rate process via:
\begin{equation}\label{diff.rate}
r_t = {d\ln(B_t)\over dt} = r^f_t - r^d_t
\end{equation}
where $r^f_t$ and $r^d_t$ are the foreign and domestic short-rate processes:
\begin{eqnarray}
 &&r_t^f = {d\ln(B^f_t)\over dt}\\
 &&r_t^d = {d\ln(B^d_t)\over dt}
\end{eqnarray}
Note, however, that $r_t$ need not be positive. Also, here we are assuming that $r_t^d$ is deterministic; however, $r_t^f$ is not, nor is $r_t$. With this assumption, using (\ref{tildeV}), we can compute the actual price ${\widetilde V}_t$ of the claim $Y_T$ by computing the would-be ``price'' $V_t$ of the claim $Y_T$ with $S_t$ and $B_t$ playing the roles of the tradable and the cash bond, respectively. In the following, for the sake of notational and terminological convenience and brevity,\footnote{\, Alternatively, we can set $r_t^d$ to zero, so $B_t$ is the same as the foreign cash bond $B_t^f$, and restore the (deterministic) $r_t^d$ dependence at the end by multiplying all derivative prices by $B_t^d(B_T^d)^{-1}$.} we refer to $B_t$ as the cash bond, $r_t$ as the short-rate, and $V_t$ as the claim price; also, we refer to the FX rate $S_t$ as FXR.

\subsection{Boundaries}

{}The boundaries in the target zone make things trickier compared with when we have an unbounded FXR process. Here we consider a target zone with {\em attainable} boundaries, i.e., the FXR process $S_t$ is confined between a lower bound $S_-$ and an upper bound $S_+$, and it can at times take these values (cf. unattainable boundaries).\footnote{\, In practice the exchange rates in target zones frequently attain the boundaries, so attainable boundaries, which we focus on here, are more realistic and appealing from this viewpoint. In fact, in some cases the FX options markets imply a future expectation that the FX rate will break the band. For academic literature on unattainable boundaries, see \cite{FXTZ}.} This requires appropriate boundary conditions to be imposed at the boundaries (see below), which in turn implies that the FXR process $S_t$ cannot be a (geometric) Brownian motion with a drift. Instead, it must be realized as a nontrivial nonlinear function of some underlying stochastic process $X_t$.

{}This can be achieved as follows. Let us consider the process (here $W_t$ is a ${\bf Q}$-Brownian motion)
\begin{equation}\label{X}
 dX_t = \sigma(X_t) dW_t + \mu(X_t) dt
\end{equation}
where $\sigma(x)$ and $\mu(x)$ have no explicit time dependence.\footnote{\, We consider time-homogeneous dynamics so the problem is analytically tractable (see below).} In fact, for our purposes here, motivated by analytical tractability, it will suffice to consider constant $\sigma(x)\equiv\sigma$. However, for now we will keep $\mu(x)$ general (but Lipschitz continuous). We will now introduce barriers for the process $X_t$ at $X_t = x_-$ and $X_t = x_+$. Below, without loss of generality, we will assume $x_- < x_+$. The FXR process is then given by $S_t = f(X_t)$, where $f(x)$ is a bounded {\em monotonic}\footnote{\, Monotonicity is assumed so there is a 1-to-1 correspondence between $S_t$ and $X_t$. Otherwise, among other things, we could not price claims, etc. (see below).} function on $[x_-,x_+]$ such that $f(x_\pm)=S_\pm$. In this regard, $S_t$ cannot have any explicit $t$ dependence,\footnote{\, Otherwise, barring any contrived time dependence, $S_t$ generically will break the band.} i.e., $S_t$ depends on $t$ only via $X_t$.

{}This has two immediate consequences. Thus, we have
\begin{equation}\label{dS}
 dS_t = f^\prime(X_t)\sigma dW_t + \left[\mu(X_t)f^\prime(X_t) + {\sigma^2\over 2}f^{\prime\prime}(X_t)\right]dt
\end{equation}
where $f^\prime(x) = \partial_x f(x)$. On the other hand, recall that $Z_t$ in (\ref{ZX}) is a martingale under the risk-neutral measure ${\bf Q}$. Taking into account (\ref{diff.rate}), it then follows that the differential rate $r_t$ cannot be arbitrary but must be a deterministic function of $X_t$ (and it has no explicit time dependence, only through $X_t$) given by $r_t = r(X_t$), where
\begin{equation}\label{r}
 r(x) = \mu(x) {f^\prime(x)\over f(x)} + {\sigma^2\over 2} {f^{\prime\prime}(x)\over f(x)}
\end{equation}
Furthermore, we must have reflecting (Neumann) boundary conditions for $f(x)$ at $x_\pm$:
\begin{equation}\label{Neumann}
 f^\prime(x_\pm) = 0
\end{equation}
Indeed, intuitively, with any other boundary conditions (absorbing, that is, Dirichlet; or Robin, i.e., mixed) the volatility $f^\prime(X_t)\sigma$ in (\ref{dS}) would be nonvanishing at the boundaries and this would imply that the process $S_t$ cannot stay in the target zone.\footnote{\, Recall that $dW_t$ is of order $\sqrt{dt}$, so, unless the volatility $f(X_t)\sigma$ vanishes at the boundaries, no drift term can compensate the stochastic term in (\ref{dS}) breaking through the barriers.} Another way of phrasing this is that, unless we have the reflecting boundary conditions (\ref{Neumann}), the identity process $I_t\equiv 1$ cannot be a martingale and probability would invariably leak through the barriers \cite{FXTZ}.

{}Let us note that (\ref{r}) is a consequence of Uncovered Interest Rate Parity (UIRP) (see fn. \ref{fn.UIRP}). Indeed, $Z_t$ defined in (\ref{ZX}) being a martingale is equivalent to stating that the instantaneous rate of return of ${\widetilde S}_t$ is the same as that of the foreign cash bond $B_t^f$. On the other hand, ${\widetilde S}_t$, which is given by (\ref{S-tilde}), is the worth of the domestic cash bond in the foreign currency. I.e., $Z_t$ being a martingale, which implies (\ref{r}), is the statement that the domestic and foreign cash bonds have the same rate of return (in foreign currency), which is nothing but UIRP.

\subsection{``Pegged'' Interest Rate}

{}So, (at least in theory) the foreign monetary authority (HKMA in the case of HKD) ``pegs'' the foreign interest rates to the domestic ones to reflect the premium/discount between the exchange rate $S_t$ w.r.t. a reference rate, call it $S_*$. Assuming that the band is symmetrical is not critical here, but we will do so in the following to simplify the discussion. We can set $x_\pm = \pm L$, $f(0) = S_*$, and $S_+ - S_* = S_* - S_-$. Also, let
\begin{equation}\label{h}
 f(x) = S_*\left[1 + \gamma h(x)\right]
\end{equation}
where $h(-x) = -h(x)$ takes values in $[1,-1]$ (i.e., $h(\pm L) = \pm 1$), and $\gamma = |S_\pm/S_* - 1| \ll 1$ is the relative half-width of the band.

{}The ideal interest rate peg then is given by $r^f_t = r^d_t + r_t$, where $r_t = r(X_t)$ and the function $r(x)$ is given by (\ref{r}). However, $X_t$ is not an observable process, only $S_t = f(X_t)$ is.\footnote{\, Albeit, since $f(x)$ is a monotonic function on $[x_-,x_+]$, there is a 1-to-1 correspondence between $X_t$ and $S_t$. Without this, the peg would not be possible, nor would it be possible to price options, e.g., calls and puts on FXR \cite{FXTZ}.} So, in practice the foreign monetary authority would have to choose some function $\rho$ of the observable FXR process $S_t$, i.e., $r_t = \rho(S_t)$. Then (\ref{r}) becomes a differential equation for $f(x)$:
\begin{equation}\label{f}
 {\sigma^2\over 2} {f^{\prime\prime}(x)\over f(x)} + \mu(x) {f^\prime(x)\over f(x)} - \rho(f(x)) = 0
\end{equation}
If we stick to the symmetrical target zone (\ref{h}), then, to the leading order in $\gamma$, we can assume that $\rho(S_* + y) = -\rho(S_* - y)$. Also, consistently with UIRP, $\rho(S_+) < 0$ and $\rho(S_-) > 0$: indeed, when the foreign currency is weak (strong), the foreign interest rate must be lower (higher) than the domestic one.

{}A simple choice\footnote{\, The explicit form of $\rho(y)$ is not critical to our discussion here.} is given by
\begin{equation}\label{rho}
 \rho(y) = r_*\left[1 - {y\over S_*}\right]
\end{equation}
where $r_*$ is a constant to be determined. The differential equation (\ref{f}) then takes the following form:
\begin{equation}\label{f1}
 {\sigma^2\over 2} f^{\prime\prime}(x) + \mu(x) f^\prime(x) - r_* f(x) \left[1 - {f(x)\over S_*}\right] = 0
\end{equation}
Thus, e.g., for vanishing drift $\mu(x)\equiv 0$, (\ref{f1}) simplifies to
\begin{equation}\label{f2}
 {\sigma^2\over 2} f^{\prime\prime}(x) - r_* f(x) \left[1 - {f(x)\over S_*}\right] = 0
\end{equation}
This equation can be solved in terms of Jacobi elliptic functions. Thus, the equation
\begin{equation}
 \partial_u^2 \phi = 2 - 4(1+m)\phi + 6m\phi^2
\end{equation}
is solved by $\phi(u) = \mbox{sn}^2(u|m)$, where $\mbox{sn}(u|m)$ is the Jacobi elliptic function, and $m$ is the elliptic parameter.\footnote{\, See, e.g., \cite{AbSt1964}, \cite{Carlson2006}, \cite{DFML}, \cite{Johannessen2018}.} However, precisely because the target zone is narrow, i.e., because $\gamma \ll 1$, the solution is actually well-approximated by elementary functions.

\subsection{Vanishing Drift}

{}In the case of vanishing drift $\mu(x) \equiv 0$, the dynamics of the process $X_t$ inside the band is purely stochastic. This can be interpreted, among other things, as the absence of any intervention from the foreign monetary authority, i.e., no monetary policy is applied to the exchange rate inside the band (which is what one expects to be the case, at least in theory -- see below). We can solve (\ref{f2}) as follows. Note that $\gamma\ll 1$ in (\ref{h}). Then, to the leading order in $\gamma$, we have
\begin{equation}
 {\sigma^2\over 2} h^{\prime\prime}(x) + r_* h = 0
\end{equation}
subject to the Neumann boundary conditions (\ref{Neumann}) (also, recall that $h(x) = -h(-x)$, and we normalize $h(x)$ such that $h(\pm L) = \pm 1$):
\begin{equation}
 h^\prime(\pm L) = 0
\end{equation}
The solution is given by
\begin{eqnarray}\label{h.sin}
 &&h(x) = \sin\left({\pi x \over 2L}\right)\\
 &&r(x) = -\gamma r_* h(x)\\
 &&r_* = r_0\label{r_0}\\
 &&r_0 = {\pi^2\sigma^2\over 8L^2}
\end{eqnarray}
So, the differential rate must take (positive and negative) values with the absolute value of order $\gamma r_*$. To put this into perspective, we must pick a time horizon $T$, which we take to be 1 year. In the case of USD/HKD, we have $\gamma = 0.05 / 7.80$, so
\begin{equation}\label{swing}
 \gamma r_* \approx 0.79\% \times {\sigma^2\over L^2}
\end{equation}
If the volatility were such that $\sigma\sqrt{T} \sim L$, then the foreign interest rate swings (\ref{swing}) due to the peg would be within a reasonable range. However, in practice FXR can swing in the target zone from one boundary to another many times in 1 year, so we actually have $\sigma\sqrt{T} = \beta L$, where the factor $\beta \sim \sqrt{K}$ roughly accounts for the number $K$ of such swings per annum and generally we can have $\beta \gg 1$ (if $K$ is large).\footnote{\, Thus, consider a simple discrete-time model, where time $t=1, 2, \dots, T$ is measured in days, and $X_t$ can only take two values, $+L$ and $-L$ (albeit $X_{t+1}$ need not be equal $-X_t$, it can also be equal $X_t$). Assuming no drift (i.e., the average of $X_t$ for large $T$ is zero), the daily variance $\sigma^2$ is given by $KL^2/(T-1)$, where $K$ is the number of times $X_{t+1} = -X_t$ (so, then $T-1-K$ is the number of times $X_{t+1} = X_t$). So, the annualized volatility $\sigma\sqrt{T} \approx L\sqrt{K}$.} Then the foreign interest rate swings (\ref{swing}) could be too large ($\sim \gamma\beta^2$) to be realistic. This implies that without a drift the ``no arbitrage'' (i.e., UIRP) condition (\ref{r}) cannot be satisfied in a high-volatility environment.

\subsection{Nonvanishing Drift}

{}With a nonvanishing drift, there is more structure. Let $r(x)$ be given by (\ref{rho}). To the leading order in $\gamma$ we have
\begin{equation}\label{h.mu}
 {\sigma^2\over 2} h^{\prime\prime}(x) + \mu(x) h^\prime(x) + r_* h(x) = 0
\end{equation}
Let us define
\begin{equation}\label{psi.h}
 \psi(x) = \exp\left({1\over\sigma^2}\int_{x_-}^x dy~\mu(y)\right) h(x)
\end{equation}
The equation for $\psi(x)$ then reads:
\begin{equation}\label{psi}
 -{\sigma^2\over 2} \psi^{\prime\prime}(x) + V(x)\psi(x) = E\psi(x)
\end{equation}
where $E = r_*$ and
\begin{equation}\label{V}
 V(x) = {\mu^2(x)\over 2\sigma^2} + {\mu^\prime(x) \over 2}
\end{equation}
Also, we have the following boundary conditions (which follow from (\ref{Neumann}))
\begin{equation}\label{b.psi}
 \psi^\prime(x_\pm) = {\mu(x_\pm)\over \sigma^2} \psi(x_\pm)
\end{equation}
So, (\ref{psi}) is the Schr\"{o}dinger equation \cite{Schrodinger1926} for a particle in a 1-dimensional box in the potential $V(x)$. In fact, this potential corresponds to supersymmetric quantum mechanics (see, e.g., \cite{Bernstein1984}, \cite{Marchesoni1988}, \cite{Witten1981}), so its spectrum is nonnegative. Due to the boundary conditions (\ref{b.psi}), we have a discrete spectrum of energy levels (eigenvalues) $E = E_n$, $n=0,1,2,\dots$; the ground level has zero energy $E_0 = 0$ with the eigenfunction $\psi_0(x)$ given by ($a_0$ is a normalization constant)
\begin{equation}
 \psi_0(x) = a_0 \exp\left({1\over\sigma^2}\int_{x_-}^x dy~\mu(y)\right)
\end{equation}
Here we are interested in the first excited level with the energy $E_1$, whose eigenfunction $\psi_1(x)$ has just one node, i.e., it vanishes only for one value of $x$ in $[x_-,x_+]$ (see above). This solution and the value of $E_1$ depend on the functional form of the drift $\mu(x)$. Based on our discussion above, ideally we wish to find configurations with the lowest possible $E_1 = r_*$, as this would minimize the foreign interest rate swings.

{}At first it might seem that a mean-reverting drift is what we need to decrease $r_*$. This (misguidedly) would appear to be the case based on the observation that for a mean-reverting drift the potential $V(x)$ would be lower compared with a ``momentum'' drift. E.g., if $\mu(x) = - \alpha x$, where $\alpha > 0$, then $X_t$ is a mean-reverting Ornstein-Uhlenbeck (OU) process \cite{OU} (with the long-run mean equal 0), and the potential is given by $V(x) = {1\over 2}\left[-\alpha + \alpha^2x^2/\sigma^2\right]$. Its ``momentum'' counterpart drift $\mu(x) = \alpha x$ has a higher potential $V(x) = {1\over 2}\left[\alpha + \alpha^2x^2/\sigma^2\right]$, which na\"{i}vely might appear to result in higher $r_* = E_1$.

{}However, the above reasoning is flawed. This is because $E_1$ is not only sensitive to $V(x)$, but also (and more so -- see below) to the boundary conditions (\ref{b.psi}), which are affected by the choice of $\mu(x)$. To understand the dependence of $E_1$ on $\mu(x)$, let us take the above linear drift $\mu(x) = \epsilon\alpha x$ (where $\alpha > 0$, and $\epsilon=-1$ corresponds to mean-reversion, while $\epsilon=+1$ corresponds to ``momentum'')\footnote{\, We use ``momentum'' in quotation marks as the drift $\mu(x)$ has nontrivial $x$-dependence.} and assume that $\alpha$ is small, that is, $\alpha \ll r_0$. Then the first term in the potential (\ref{V}) is subleading and can be neglected, i.e., the potential is approximately constant. The desired solution is given by ($a_1$ is a normalization constant)
\begin{eqnarray}
 &&\psi_1(x) = a_1\sin(\omega x)\\
 &&E_1 = {1\over 2}\left[\epsilon\alpha + \sigma^2\omega^2\right]
\end{eqnarray}
Now, $\omega$ is determined from the boundary conditions (\ref{b.psi}). Straightforward algebra gives (to the first order in $\alpha/r_0$)
\begin{equation}
 \omega = {\pi\over 2L} - {2\epsilon\alpha L\over\pi\sigma^2}
\end{equation}
So, we have (again, to the first order in $\alpha/r_0$)
\begin{equation}
 E_1 = r_0 - {\epsilon\alpha\over 2}
\end{equation}
So, the mean-reverting drift ($\epsilon = -1$) increases $E_1$, while it is the ``momentum'' drift ($\epsilon = +1$) that decreases it. This is because the increase (decrease) in $\omega$ for $\epsilon=-1$ ($\epsilon=+1$) due to the boundary conditions (\ref{b.psi}) overweighs the decrease (increase) in the potential. We arrived at this conclusion for small $\alpha$. When $\alpha$ is not small, the solution to (\ref{psi}) involves parabolic cylinder functions and $E_1$ can be computed numerically as a function of $\alpha$. However, on general grounds we expect that, as $\alpha$ increases, the ratio $E_1/r_0$ is reduced (for $\epsilon=+1$) down from 1 by some factor $\kappa \sim 1$.

{}This can be seen explicitly by considering simpler drifts. Thus, let us take $\mu(x) = \epsilon\nu\sigma^2\mbox{sign}(x)$, where $\nu > 0$ is a constant. So, we have a constant drift on each side of $x=0$. (The discontinuity in the drift does not pose a problem. We will consider smooth drifts below.) As above, for $\epsilon=-1$ ($\epsilon = +1$) we have mean-reversion (``momentum''). The potential is given by $V(x) = {\sigma^2\over 2}\left[2\epsilon\nu\delta(x) + \nu^2\right]$. So, it is a constant potential augmented with a delta-function spike at $x = 0$. The delta-function affects even levels $E_n$, $n=0,2,4,\dots$; however, it does not affect the odd levels $E_n$, $n=1,3,\dots$, as for such levels we have $\psi_n(0) = 0$. So, the solution for $\psi_1(x)$ is given by ($a_1$ is a normalization constant)
\begin{eqnarray}\label{psi.sign}
 &&\psi_1(x) = a_1\sin(\omega x)\\
 &&E_1 = {\sigma^2\over 2}\left[\nu^2 + \omega^2\right]\label{E1.sign}
\end{eqnarray}
As above, $\omega$ is determined from the boundary conditions (\ref{b.psi}):
\begin{equation}\label{b.eq.sign}
 \omega\cot(\omega L) = \epsilon\nu
\end{equation}
Then for $\epsilon=-1$ we have $\omega > \pi/2L$, while for $\epsilon=+1$ we have $\omega < \pi/2L$, i.e., a mean-reversion (``momentum'') drift increases (decreases) $\omega$. For small $\nu \ll 1/L$ we have (to the first order in $\nu$)
\begin{eqnarray}
 &&\omega = {\pi\over 2L} - {2\epsilon\nu\over \pi}\\
 &&E_1 = r_0\left[1 - {8\epsilon\nu L\over \pi^2}\right]
\end{eqnarray}
So, as in the linear drift case above, mean-reversion (``momentum'') increases (decreases) $E_1$. We can solve (\ref{b.eq.sign}) numerically for values of $\nu$ that are not small. For $\epsilon = -1$ we have $E_1 > r_0$. For $\epsilon = +1$, as $\nu$ increases, $E_1$ decreases until the limiting value $\nu = 1/L$, for which the solution is actually linear (and is obtained from (\ref{psi.sign}) in the limit $\omega \rightarrow 0$, $a_1\rightarrow\infty$, ${\widetilde a}_1=a_1\omega = \mbox{finite}$):
\begin{eqnarray}\label{psi.sign.1}
 &&\psi_1(x) = {\widetilde a}_1 x\\
 &&E_1 = {4r_0\over\pi^2}\approx 0.41 \times r_0\label{min.E1.sign}
\end{eqnarray}
For $\nu > 1/L$ (\ref{b.eq.sign}) has no roots ($\epsilon=+1$), so (\ref{min.E1.sign}) is the lowest value for this drift.

{}As mentioned above, the $\mbox{sign}(x)$ discontinuity in the drift does not pose a problem. However, we can consider a smooth ``momentum'' drift of the form $\mu(x) = \nu\sigma^2\tanh(\nu x)$, for which we have a constant potential $V(x) = {1\over 2}\sigma^2\nu^2$. So, the solution is still given by (\ref{psi.sign}) and (\ref{E1.sign}), but the boundary conditions (\ref{b.psi}) now read
\begin{equation}\label{b.eq.tanh}
 \omega\cot(\omega L) = \nu\tanh(\nu L)
\end{equation}
For small $\nu \ll 1/L$ we have (to the first order in $\nu^2$)
\begin{eqnarray}
 &&\omega = {\pi\over 2L} - {2\nu^2 L\over \pi}\\
 &&E_1 = r_0\left[1 - {8\epsilon\nu^2 L^2\over \pi^2}\right]
\end{eqnarray}
We can solve (\ref{b.eq.tanh}) numerically for values of $\nu$ that are not small. As above, the lowest $E_1$ corresponds to the aforementioned limit $\omega \rightarrow 0$ (where we have the linear solution (\ref{psi.sign.1})), which occurs when $\nu L\tanh(\nu L) = 1$ at $\nu\approx 1.20/L$:
\begin{equation}
 E_1 \approx 0.58 \times r_0
\end{equation}
For completeness, let us also note that the mean-reversion analog of the ``momentum'' drift $\mu(x) = \nu\sigma^2\tanh(\nu x)$ is given by $\mu(x) = -\nu\sigma^2\tan(\nu x)$, where $\nu > 0$. For this drift we invariably have $E_1 > r_0$.

{}So, while na\"{i}vely it might have seemed that mean-reverting drifts would aid in lowering $r_*$, they have exactly the opposite effect, and it is in fact ``momentum'' drifts that do the job. An intuitive way of thinking about this is as follows. Mean-reverting drifts actually increase the effective volatility as the process $X_t$ bounces back and forth between the boundaries more (compared with the vanishing drift case), i.e., they effectively increase the factor $\beta\sim\sqrt{K}$ (see above). On the other hand, ``momentum'' drifts push $X_t$ toward the boundaries and make it spend more time thereat, thereby effectively decreasing $\beta\sim\sqrt{K}$. Furthermore, in high-volatility environments a realistic value of $r_*$ based on the foreign interest rate policy will have sizably lower implied volatility than that required by the no-arbitrage condition (\ref{r}). That is, realistically, in high-volatility environments there should exist arbitrage opportunities.\footnote{\, By ``arbitrage'' we do not mean risk-free arbitrage. Deviations from UIRP are not risk-free arbitrage opportunities (see above).} However, these are not necessarily vanilla FX carry trade opportunities: indeed, if, as in (\ref{rho}), $r_t = r_*\left[1 - S_t / S_*\right]$, then there is no vanilla FX carry trade (ignoring any transaction costs, tax considerations, etc.), where one would borrow (lend) the domestic currency and lend (borrow) the foreign currency and hold a static position. This is because the differential rate $r_t$ is pegged to FXR -- albeit possibly more weakly than the no-arbitrage condition (\ref{r}) would require -- and its sign can flip. Instead, the arbitrage opportunity here is dynamic.

{}To understand this, let us assume that $f(x)$ is such that it satisfies (\ref{f1}) with some $r_*$. Let us further assume that the foreign monetary authority pegs the foreign interest rate such that $r_t = r_{**}\left[1 - S_t/S_*\right]$, where $r_{**} < r_*$, in fact, $r_{**}$ could even be sizably smaller than $r_*$ (see above). Recall that the foreign tradable ${\widetilde S}_t = B_t^d S_t$ is the value of the domestic bond in the foreign currency. Assuming (\ref{f1}), the instantaneous rate of return of ${\widetilde S}_t$ is given by ${\widetilde r}_t = r_t^d + r_*\left[1 - S_t/S_*\right]$. Since by definition $r_t = r_t^f - r_t^d$, it then follows that ${\widetilde r}_t > r_t^f$ when $S_t < S_*$, and ${\widetilde r}_t < r_t^f$ when $S_t > S_*$. So, the arbitrage strategy (again, ignoring transaction costs, taxes, etc.) then consists of being long the domestic bond and short the foreign bond when $S_t < S_*$, and being short the domestic bond and long the foreign bond when $S_t > S_*$. So, this is a dynamic combination of two opposite carry trades depending on where FXR is w.r.t. $S_*$. However, the existence of this arbitrage opportunity (which may or may not be actionable once transaction costs, tax considerations, etc., are taken into account) does not necessarily compel the foreign monetary authority to peg the foreign interest rate with an unrealistic value of $r_*$ (if the volatility $\sigma$ is high enough, that is) as there are other more important economic considerations at stake. Thus, deviations from UIRP and the corresponding carry trades exist for some FX pairs and there have been many academic papers (for a partial list, see, e.g., the literature cited in fn. \ref{fn.UIRP} and references therein) written on the so-called ``Forward Discount Puzzle'', which nonetheless do not make it go away.

{}So, as we saw above, mean-reversion inside the band exacerbates the volatility/differential rate issue, while ``momentum'' makes it milder. This might seem puzzling at first, as it would make no sense that the foreign monetary authority would induce ``momentum''-like behavior inside the band via its monetary policy interventions. And (at least in theory) it does not and should not. Assuming FXR is allowed to float freely inside the target zone, the foreign monetary authority has no business in controlling FXR's behavior inside the band. Instead, its primary concern and responsibility is to make sure that FXR does not breach the band, which is achieved via monetary policy interventions at the {\em boundaries} of the target zone. We will discuss this below in a moment. Before we do this, however, let us mention a very simple way of reducing the volatility.

{}Consider a discrete setup, where $S_t$ can only take two values, $S_+$ and $S_-$, but the values between $S_+$ and $S_-$ are not permitted. This is basically a scenario where there is a market-maker (or multiple ones) that are willing to buy each unit of the domestic currency at $S_-$ (the bid) and sell it at $S_+$ (the ask). If the bid-ask spread is large enough, then aggressive order flow is not cheap and this will typically reduce the volatility, i.e., the process $S_t$ will spend considerable amount of time at $S_+$ or $S_-$. This may not be a very ``democratic'' setup and might be argued to defy the idea behind having free-floating FXR inside the band. However, a priori, it could be a viable scenario if reducing the volatility is the goal.

\subsection{Intervention at the Boundaries}

{}To ensure that FXR does not breach the band, the foreign monetary authority must intervene at the boundaries, i.e., when $S_t$ hits $S_+$ or $S_-$. When $S_t = S_+$, the foreign currency is weak and there is pressure for it to become weaker. To prevent this, the foreign monetary authority buys the foreign currency (i.e., reduces its supply) to stabilize it. Similarly, when $S_t = S_-$, the foreign currency is strong and there is pressure for it to become stronger. To prevent this, the foreign monetary authority sells the foreign currency (i.e., increases its supply) to stabilize it. In particular, this is how HKMA operates in the case of USD/HKD.

{}So, how can we model this monetary intervention at the boundaries? To begin with, let us note that a priori we can build an infinite number of such models by choosing the drift (see below). However, the idea behind them is the same. Instead of confining the process $X_t$ to the interval $[x_-,x_+]$ (see above), we can let $X_t$ take values on the entire real line ${\bf R}$. However, we can effectively confine the process $X_t$ to $[x_-,x_+]$ by having a large mean-reverting drift outside this interval. In fact, for our discussion here what happens inside this interval is of no import, so for the sake of simplicity we can assume vanishing drift $\mu(x) = 0$ for $x_- < x < x_+$, but it is nonvanishing outside this interval (as above, we will set $x_\pm = \pm L$):
\begin{equation}
 \mu(x) = \xi\sigma^2\left[\theta(-x - L) - \theta(x - L)\right]
\end{equation}
where $\theta(y)$ is the Heaviside step-function. Here we are going to assume that $\xi L \gg 1$, so we have a large mean-reverting drift outside the boundaries located at $x_\pm = \pm L$. The potential is given by
\begin{equation}
 V(x) = {\sigma^2\over 2}\left[-\xi\delta(x + L)-\xi\delta(x - L) + \xi^2\theta(-x - L) + \xi^2 \theta(x - L)\right]
\end{equation}
The solution to (\ref{psi}) for the first excited level $E_1=r_*$ is given by ($\psi_1(-x) = -\psi_1(x)$):
\begin{eqnarray}
 &&\psi_1(x) = a_1\sin(\omega x),~~~0\leq x \leq L\\
 &&\psi_1(x) = a_1\sin(\omega L)\exp(-\lambda(x - L)),~~~x\geq L\\
 &&\lambda = \sqrt{\xi^2 - \omega^2} \approx \xi - {\pi^2\over8\xi L^2}\\
 &&\sin(2\omega L) = {\omega\over\xi}~~~\Rightarrow~~~\omega \approx {\pi\over 2L}\left[1 - {1\over 2\xi L}\right]\\
 &&E_1 = {\sigma^2\omega^2\over 2}
\end{eqnarray}
where $a_1$ is a normalization constant. So, using (\ref{psi.h}), we have (again, recall that $h(-x) = -h(x)$):\footnote{\, This result for $h(x)$ can be obtained directly from (\ref{h.mu}) by requiring that $h(x)$ and $h^\prime(x)$ are continuous at $x = \pm L$.}
\begin{eqnarray}
 &&h(x) = {\sin(\omega x)\over\sin(\omega L)},~~~0\leq x \leq L\\
 &&h(x) = \exp\left[(\xi-\lambda)(x - L)\right] \approx \exp\left[{\pi^2(x - L)\over 8\xi L^2}\right],~~~x\geq L
\end{eqnarray}
where, as above, we have normalized $h(x)$ such that $h(\pm L) = \pm 1$. So, as desired, $h(x)$ is a monotonically increasing function of $x$; however, for $|x| > L$ it increases very slowly (recall that $\xi L \gg 1$), which is due to the large negative drift for $x > L$ and large positive drift for $x < -L$. In the limit $\xi \rightarrow \infty$, $h(x)$ flattens at $|x| > L$, so we have reflecting barriers. This neatly models the intervention by the foreign monetary authority at the boundaries. Also, note that we must have $\xi > 0$ for the above solution to exist, i.e., we must have (strong) mean-reversion outside the band.

{}The upshot is that FXR is kept inside the band via the intervention by the foreign monetary authority at the boundaries, not inside the target zone. Without such intervention there is no reason why FXR would stay inside the band. The stochastic nature of FXR invariably would lead to it breaching the band without such intervention. If, as in the USD/HKD case, the foreign currency (HKD) is fully backed by the domestic currency (USD) reserves, to keep the foreign currency inside the target zone, the foreign monetary authority (HKMA) {\em must} be bound by clear rules that govern its reaction to FXR hitting the boundaries. There can be no uncertainty about whether the foreign monetary authority will protect the band (or how it will do it), or else FXR will breach it.

\subsection{Pricing}

{}For the sake of completeness, let us discuss some aspects of derivatives pricing in the target zone. Some parts of this subsection closely follow \cite{FXTZ}, which provides additional details.

{}Recall that $r_t = r(X_t)$, where $r(x)$ is a deterministic function of $x$ (and has no explicit dependence on time $t$). Furthermore, let us focus on claims of the form $Y_T = Y(X_T)$ (see (\ref{V_t})). Let us define the pricing function
\begin{equation}
 v(x, t, T) = B_t~\mathbb{E}\left(B_T^{-1}Y_T\right)_{{\bf Q}, X_t = x}
\end{equation}
Then $V_t = v(X_t, t, T)$. Since $B_t^{-1}V_t$ is a ${\bf Q}$-martingale, we have the following PDE for $v(x,t,T)$:
\begin{equation}\label{PDE.v}
 \partial_t v(x,t,T) + \mu(x) \partial_x v(x,t,T) + {\sigma^2\over 2}\partial_x^2 v(x,t,T) - r(x) v(x,t,T) = 0
\end{equation}
subject to the terminal condition $v(x,T,T) = Y(x)$. Also, recall that $r(X_t) = dB_t/dt$. Further, note that the claim function $Y(x)$ must satisfy the Neumann boundary conditions $\partial_x Y(x_\pm) = 0$. This follows from the fact that the pricing function $v(x, t, T)$ must satisfy these boundary conditions: $\partial_x v(x, t, T) = 0$. This in turn follows from the fact that, since the function $f(x)$ describing the FXR process $S_t = f(X_t)$ is monotonically increasing, instead of viewing $v(x, t, T)$ as a function of $x$, we can view it as a function of $s = f(x)$. Thus, we have $\partial_x v = \partial_s v f^\prime(x)$, so $\partial_x v$ indeed vanishes at the boundaries as $f^\prime(x_\pm) = 0$.

{}Furthermore, we can rewrite (\ref{PDE.v}) as a differential equations w.r.t. $s$ (we have used (\ref{r}) in deriving this equation)
\begin{equation}\label{PDE.s}
 \partial_t {\widehat v}(s, t, T) + \rho(s) \left[s \partial_s {\widehat v}(s, t, T) - {\widehat v}(s, t, T)\right] + {\sigma^2\over 2}g(s) \partial_s^2 {\widehat v}(s, t, T) = 0
\end{equation}
Here ${\widehat v}(s, t, T) = v(x, t, T)$, $g(s) = [f^\prime(x)]^2$, and $\rho(s) = r(x)$, where $x$ is determined from $f(x) = s$. A significant simplification arises for a ``linear peg'' (see (\ref{rho})):
\begin{equation}\label{rho.s}
 \rho(s) = r_*\left[1 - {s\over S_*}\right]
\end{equation}
Then we have the following solution:
\begin{equation}
 {\widehat v}(s, t, T) = a{s\over S_*} + b\left[1 - {s\over S_*}\right] e^{-r_*(T-t)}
\end{equation}
where $a$ and $b$ are constants. Thus, for a claim $Y_T = S_T$, which corresponds to a forward, we have $b=0$ and $a=S_*$, so ${\widehat v}(s, t, T) = S_t$, as it should be. On the other hand, for a claim $Y_T = 1$, which corresponds to a zero-coupon $T$-bond (with the principal given by 1 unit of the foreign currency paid at maturity $T$), we have $a = 1$ and $b = 1$, so
\begin{equation}
 {\widehat v}(s, t, T) = {s\over S_*} + \left[1 - {s\over S_*}\right] e^{-r_*(T-t)}
\end{equation}
Assuming, for simplicity, that the domestic rate $r_t^d = r^d$ is constant, we have the following price for the $T$-bond (see (\ref{tildeV})):
\begin{equation}\label{bond}
 P(t, T) = {S_t\over S_*} e^{-r^d(T-t)} + \left[1 - {S_t\over S_*}\right] e^{-{\widehat r}(T-t)}
\end{equation}
where ${\widehat r} = r_d + r_*$. Assuming that $r_* < r^d / \gamma$, straightforward algebra gives that the r.h.s. in (\ref{bond}) is always less than 1, as it should be. Indeed, the foreign short rate process is given by $r_t^f = r_t^d + r_*\left[1-S_t/S_*\right]$, and it is always positive assuming that $r^d > \gamma r_*$. Then the $T$-bond price, which is given by
\begin{equation}
 P(t, T) = \mathbb{E}\left(\exp\left[-\int_t^T dt^\prime r^f_{t^\prime}\right]\right)_{{\bf Q}, {\cal F}_t}
\end{equation}
is always less than 1. On the other hand, as we discuss above, in high-volatility environments $r_*$ dictated by the no-arbitrage condition (\ref{r}) can be too large for realistic values of $r^d$, so $r^f_t$ would become negative and the $T$-bond price could become greater than 1. On the flipside it would also mean that the maximum value of $r_t^f = r^d + \gamma r_*$ would be unrealistically high. So, in such real-life scenarios the foreign monetary authority cannot peg the foreign interest rate to satisfy (\ref{r}) and (as we discuss in the main text of this paper) the band will be breached without the foreign monetary authority's intervention at the boundaries. I.e., without the foreign monetary authority's intervention, in high-volatility environments the foreign interest rate $r^f_t$ would have to be unrealistically high when $S_t$ approaches $S_-$ (strong side), and it would have to become negative when $S_t$ approaches $S_+$ (weak side).

{}At first, there might appear to be a ``simple'' solution that would prevent $r^f_t$ from turning negative. Thus, instead of $r_t = r_*\left[1-S_t/S_*\right]$, we can attempt to have a lopsided differential rate $r_t = r_*\left[1-S_t/S_+\right]$, so $r_t$ never turns negative, and, consequently,\footnote{\, Strictly speaking, we can have $r_t = r_*\left[1-S_t/S_1\right]$, where $S_1$ is such that $r^d + r_*\left[1-S_+/S_1\right]\geq 0$ (here, as above, for simplicity we assume constant $r^d_t = r^d$). However, this is not critical here.} neither does $r_t^f$. However, this is trickier than it might appear. At the boundaries we must have $f^\prime(x_\pm)=0$. Looking at (\ref{r}), this then implies that we have $f^{\prime\prime}(x_-) > 0$, and $f^{\prime\prime}(x_+) = 0$. This is not possible for a ``linear peg'', i.e., the function $\rho(s)$ in (\ref{rho.s}) must be nonlinear in $s$ (recall that $r_t = \rho(S_t)$). An example would be that $S_+ - f(x) \sim (x_+ - x)^3$ as $x\rightarrow x_+$, in which case $\rho(s) \sim (S_+ - s)^{1/3}$ as $s\rightarrow S_+$. However, even if the weak-side issue (with the foreign interest rate turning negative) is solved with a contrived ``nonlinear peg'', in high-volatility environments unrealistically high interest rates would still be required at the strong-side (i.e., as $S_t$ approaches $S_-$), and the only way to ensure that the band is not breached is through interventions by the foreign monetary authority.

\end{document}